\documentclass{kapproc} 
        \usepackage{graphicx}
\let\footnote\savefootnote
\let\footnotetext\savefootnotetext 
\let\footnoterule\savefootnoterule 
\kluwerbib

\begin{document}

\articletitle{Luminosities of AGB Variables
}

\author{Patricia Whitelock}
\affil{SAAO, PO Box 9, Observatory, 7935, South Africa.}
\email{paw@saao.ac.za}

\begin{abstract}
 The prevailing evidence suggests that most large-amplitude AGB variables
follow the period luminosity (PL) relation that has been established for
Miras in the LMC and galactic globular clusters. Hipparcos observations
indicate that most Miras in the solar neighbourhood are consistent with such
a relation. There are two groups of stars with luminosities that are
apparently greater than the PL relation would predict: (1) in the LMC and
SMC there are large amplitude variables, with long periods, $P> 420$ days,
which are probably undergoing hot bottom burning, but which are very clearly
more luminous than the PL relation (these are visually bright and are likely
to be among the first stars discovered in more distant intermediate age
populations); (2) in the solar neighbourhood there are short period,
$P<235$ days, red stars which are probably more luminous than the PL
relation. Similar short-period red stars, with high luminosities, have not
been identified in the Magellanic Clouds.
 \end{abstract}


\section*{Introduction}
This review concentrates on the luminosities of large amplitude Mira-like
variables. These are of particular interest because it is during this
large-amplitude phase that most of the mass loss occurs. Furthermore, the
kinematics of these stars depend upon their pulsation period; thus, if we
can measure the period, we can tell a great deal about the star and its
parent population (Feast \& Whitelock 2000a). I discuss both O- and C-rich
variables, but concentrate on the O-rich ones about which we know most.
Because of its importance to luminosities, some emphasis is put on the
period luminosity (PL) relation, drawing on observations of globular
clusters, the Magellanic Clouds, other local group galaxies and last but
not least, the solar neighbourhood.

Before examining what we know about luminosities it is worth emphasizing the
importance of AGB variables in the understanding of extragalactic
populations. The most luminous stars present in old or intermediate age
populations are the large-amplitude AGB variables. Thus, as we become able
to resolve individual stars in ever more distant stellar populations, those
we see first and best will be this type of AGB variable. So, if we are to
use such stars as probes of their parent population, it is crucial that we
understand how their properties depend on age, metallicity etc.

\section{Globular Cluster Miras}
 The Miras in globular clusters have always been key to calibrating the
luminosity of the tip of the AGB. Unfortunately, because of their short
lifetimes there are rather few Miras in globular clusters and fewer still
that have been well studied. Let me remind you that the Miras are the most
luminous stars found in the clusters; in fact they are the only stars with
luminosities above the tip of the red giant branch (RGB). They are only
found in metal-rich clusters ($\rm [Fe/H]>-1$), and we presume that the AGB
in metal-deficient systems terminates below the tip of the RGB. The
pulsation period of a Mira is a function of the metallicity of its parent
cluster (e.g.\ Feast \& Whitelock 2000b). In fact it is only for the Miras
in clusters, and a very few in binary systems, that we can determine
metallicities. The Miras in galactic globular clusters are all O-rich and
there is no particular evidence to suggest that they have reached the
thermally pulsing part of the AGB.

Feast et al.\ (2002) recently reexamined the luminosities of the Miras in
globular clusters using a new distance calibration based on Hipparcos
parallaxes of sub-dwarfs and published photometry for 6 galactic globular
clusters, together with new observations of NGC\,121~v1, a short period low
metallicity Mira in the SMC.  They demonstrated that these cluster Miras fit
the same PL relation as do the LMC Miras, and derived a zero-point for the
PL relation. There are many globular clusters,
particularly near the galactic centre, which have not yet been properly
surveyed for Miras. We are in the process of rectifying this situation using
the Infrared Survey Facility in South Africa in collaboration with
astronomers from the University of Tokyo. Any new cluster Miras will obvious
improve our statistics, but we are particularly hopeful about finding some
longer period stars in the metal-rich bulge clusters. 

Once there are theoretical models which deal effectively with mass-loss, and
allow us to predict accurately the AGB tip luminosity for different
populations, we will be able to use them to calibrate extragalactic systems.
But, until that level of theoretical understanding is reached, those who
wish to study extragalactic systems will make deductions based on a
comparison with globular clusters or with the galactic bulge. The recent
literature contains numerous studies of AGB populations in local group
galaxies and beyond and it is interesting to see comparisons being made with
galactic globular clusters and very different conclusions being drawn by
different authors from essentially the same data. Figure 1 shows a colour
magnitude diagram for cluster variables, of the kind typically used for
comparison with extragalactic systems. The most luminous cluster star
illustrated here is NGC\,6553~v4, for which the reddening is uncertain; it
is plotted in the figure as two connected open circles for two different
reddenings.  The lower mean luminosity seems more likely and this is one
magnitude fainter than the $K=-8.5$ which some authors use.

\begin{figure}
\hspace*{2cm}\includegraphics[height=6cm]{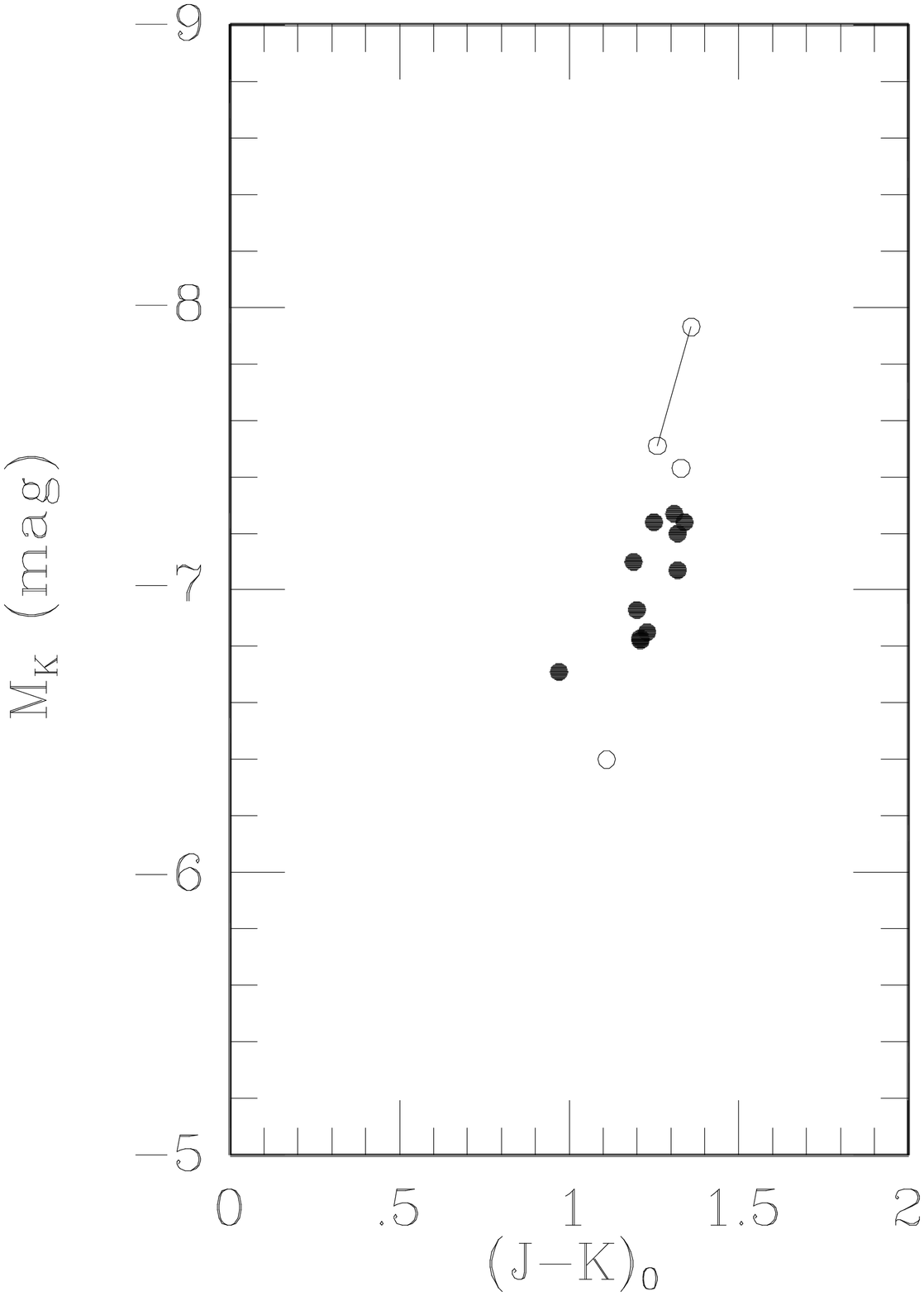} 
\vspace*{-5.5cm}
\narrowcaption{ A colour-magnitude diagram for the Miras in globular
clusters; open circles represent less certain luminosities. A luminosity
of $K= -8.5$ is often assumed for the brightest stars in globular clusters,
but $K=-7.5$ is actually a better estimate (see Feast et al. 2002).}
\vspace*{0.5cm}
\end{figure}

It is also worth noting that comparing the luminosity of individual stars in
extragalactic systems with those of AGB variables in the galactic bulge is
even more fraught with uncertainty, because the shape of the bulge and the
presence of significant numbers of foreground stars result in many stars
having distances less than that of the centre and therefore luminosities
that appear to be {\it much} brighter than they really are. 

Variability is also a factor in comparing one system with another. The short
period Miras, found in globular clusters, typically have peak-to-peak
$K$ amplitudes of around half a magnitude, so there is a high level of
uncertainty associated with single measurements of the luminosity.
Longer period stars have larger amplitudes, reaching over two magnitudes for
the 1000 day variables discussed below.

\section{Large Magellanic Cloud (LMC)}
 The PL relation for Mira variables was discovered for stars in the LMC, and
Feast et al.\ (1989) refined earlier results to show that at $K$ the O- and
C-rich stars obeyed the same same PL relation. The bolometric luminosities
seemed to show slightly different relations for the O- and C-rich stars,
although there was always the suspicion that this was an artifact of the way
that the bolometric magnitudes were calculated. Feast et al.\ also noted that
O-rich stars with $P>420$ days were significantly more luminous
than the PL relation would predict.

Working independently, but at roughly the same time, Hughes \& Wood (1990)
came to similar conclusions; although they described the PL relation as
having two linear parts, with a steeper slope over the long period ($P>400$
days) range. Their derivation of a PL relation had considerable scatter,
because their bolometric magnitudes were calculated from single
observations rather than from the mean values.

More recently we have been working on much longer period stars in the LMC,
which were discovered via their IRAS emission (e.g. Wood et al.\ 1992;
Zijlstra et al.\ 1996). Most of these are obscured stars with high mass-loss
rates. In globular cluster Miras the energy distribution peaks at a
wavelength between 1 and 2 $\mu$m, while for these IRAS sources the energy
peaks at longer wavelengths, $\lambda > 4 \mu$m. There has, as yet, been
little opportunity for systematic monitoring around the pulsation cycle at
long wavelength, although a few repeated ISO observations allow us to
estimate the bolometric amplitudes at around one magnitude. There are small
systematic differences in the bolometric magnitudes obtained using ISO and
IRAS. Furthermore, the way the colour corrections are treated can give rise
to systematic differences between O- and C-rich stars. Thus we cannot yet
claim to have accurate mean luminosities. Nevertheless, the overall
impression given by all the available results is that most of the dusty
Miras, with long periods, $420<P<1300$ days, fall on an extrapolation of the
PL relation derived for O-rich stars with $P<420$ days.

Looking in detail at the stars with bolometric magnitudes brighter than the
PL extrapolation, we find that all those which have been studied show
evidence for hot bottom burning (HBB), in particular they have very strong
lithium lines (but see also Trams et al.\ 1999). Towards the end of the AGB
evolution, in stars with initial masses in the range 4 to 6 $M_{\odot}$, the
base of the H-rich convective-envelope can dip into the H-burning shell; the
introduction of fresh H-rich material into the nuclear-burning shell allows
the luminosity to go above the core-mass luminosity predictions (Bl\"ocker
\& Sch\"onberner 1991). Carbon is burned to nitrogen, and lithium can reach
the surface via the beryllium transport mechanism (Sackmann \& Boothroyd 1992). 
Smith et al.\ (1995) surveyed luminous AGB stars in the LMC and SMC for
lithium - the clearest indication that HBB is taking place. Almost all stars
with high lithium abundance lie above the PL relation, and we can now
understand the change in slope of the PL, at around 400 to 420 days, as the
effect of HBB in LPVs {\it without} particularly thick dust shells. Those
stars with thick shells, like the ones for which we have ISO observations,
are lower mass objects, lie near the PL, and probably never experienced
HBB.

An important contribution is provided by the work of Nishida et al.\ (2000)
who monitored 3 C-rich Miras, with thick dust shells, in SMC and LMC
clusters, and were thus able to estimate pulsation periods and bolometric
magnitudes for stars with known initial mass and metallicity. The periods
are all around 500 days and the luminosities are very close to an
extrapolation of the Feast et al.\ (1989) PL for short period O-rich stars.
The turnoff masses for these clusters are around $1.5 M_{\odot}$, and
therefore much too low for their AGB stars to have undergone HBB.

\section{Other Local Group Galaxies}
 Leaving the Magellanic Clouds, but staying in the local group we look at
C-rich Miras in the Sagittarius dwarf spheroidal. This is the galaxy,
discovered only in 1994, that is merging with the Milky Way on the
far side of the bulge. It contains Miras with periods in the range 230 to
360 days, distinctly shorter than those in Magellanic Cloud clusters, as we
might expect from this somewhat older population (Whitelock et al.\ 1999). The
distance modulus derived from the Mira PL, $(m-M)_0=17.36\pm0.2$ mag, is in
good agreement with that from RR Lyrae variables, $(m-M)_0=17.18\pm0.2$ mag.

There is a Mira in IC\,1613 with a period of 641 days, the luminosity of
which is considerably brighter than the prediction of the PL relation
(Kurtev et al.\ 2001). With a spectral type of M3e it is clearly O-rich, and from 
what has been said above we must predict that it is undergoing HBB burning.
It would be interesting to look for lithium in its spectrum.

The Leo I dwarf spheroidal contains several large amplitude variables (see
Menzies, these proceedings). There is also evidence of very red C-stars from
2MASS observations of Fornax; these will almost certainly turn out to be
C-Miras. We can look forward to similar discoveries in other local group
galaxies and beyond, during the next few years.
\begin{figure}
\includegraphics[height=5cm]{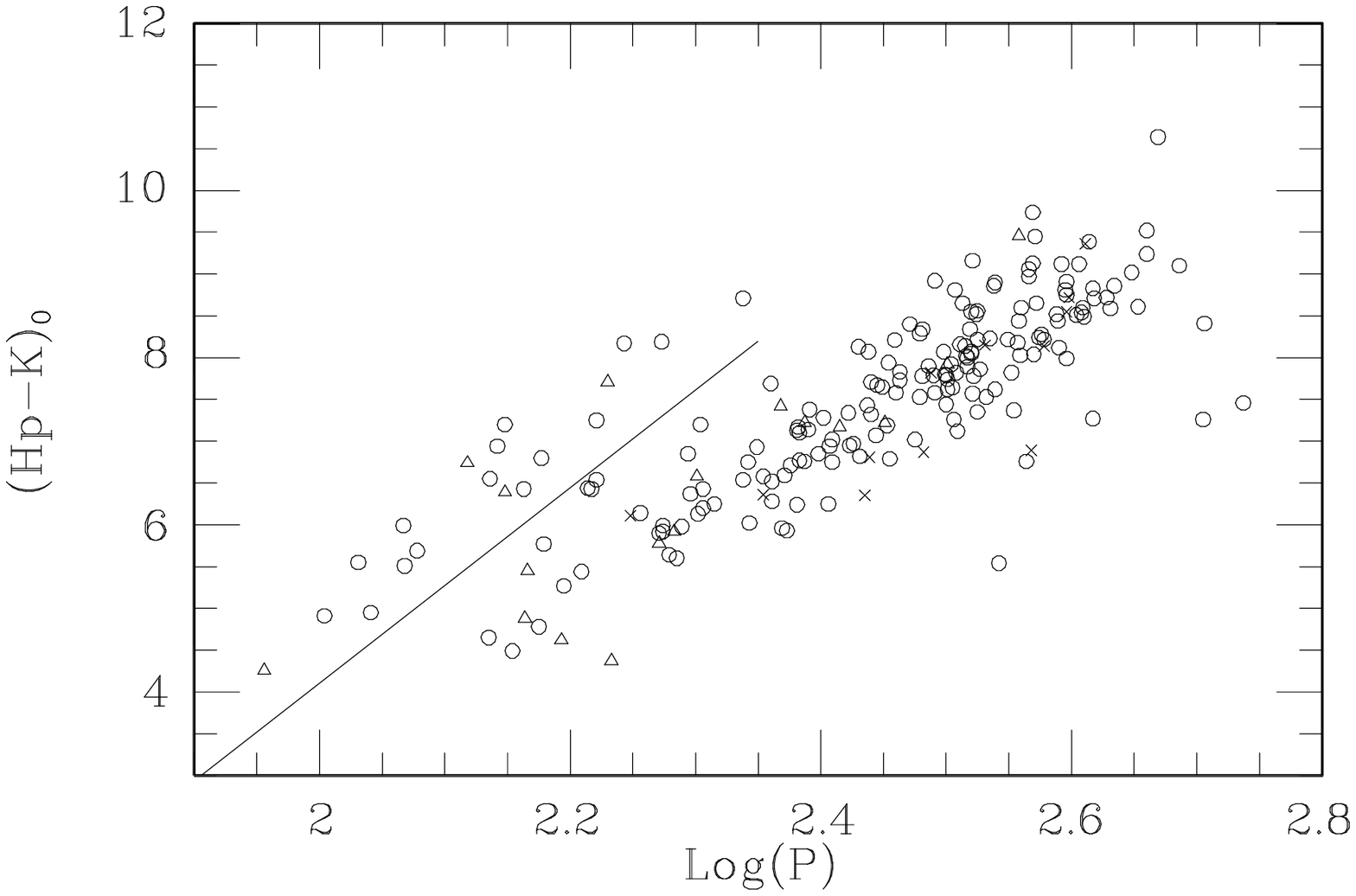}
\vspace*{-4.7cm}
\narrowcaption{A combined Hipparcos, $Hp$, near-infrared, $K$, period-colour
relation; circles represents Miras and triangles SRs, while Miras with
S-type spectra are shown as crosses; the line separates SP-red and SP-blue
stars (see Whitelock et al.\ 2000).}
\end{figure}
\section{Solar Neighbourhood} 
 Returning closer to home I want to finish by looking at what we know about
luminosities from Hipparcos parallaxes (Whitelock et al.\ 2000; Whitelock \&
Feast 2000; Feast \& Whitelock 2000a). It is worth noting that the stars
selected for the Hipparcos input-catalogue had to be visually bright
throughout their pulsation cycle. Therefore, this selection of stars have,
of necessity, low mass-loss rates, $<10^{-7} M_{\odot}yr^{-1}$; they are
different from the LMC sample discussed above. The selection
comprised 213 O-rich Mira-like variables with $K$ magnitudes and pulsation
periods.

Before discussing the parallax analysis I outline some of the
characteristics of these stars, and their dependence on period and colour.
Figure 2 shows the stars in a period-colour plot, where
$Hp$ is the Hipparcos broad-band magnitude. At short periods the variables
divide into two sequences, a blue one that contains most of the stars and a
parallel red one which contains a significant fraction of the shortest
period stars; notice that both sequences contain SRs and Miras. The straight
line in Fig~2 divides the stars into two groups: those above the line are
the short period red (or SP-red) group, while those below, which form part
of the sequence seen at longer periods, are the short-period blue (or
SP-blue) group (Whitelock et al.\ 2000).

The pulsation amplitudes of the two groups are very similar, but the colour
differences extend to their near-infrared colours - the mean spectral type
of the red group, M4.5, is later than that of the SP-blues, M3. Furthermore,
the kinematics and scale heights of the two groups are different. The
SP-blue stars have larger scale heights, a greater velocity dispersion and a
larger asymmetric drift than the SP-red stars. The SP-blue stars have
similar characteristics to the Miras found in globular clusters, while the
SP-reds seem to be rather different and apparently somewhat younger (see
also Feast these proceedings). The differences between these two groups
justifies treating them differently in the parallax analysis.

Given the high uncertainties on the parallax measurements it is not
practical to fit a PL relation to the data. Rather we assume the slope of
the PL relation ($M_K=-3.47 \log P + \beta$), derived from the LMC work, and
deduce $\beta$, the zero point, from the parallaxes. The PL relation is
solved in the form: $$ 10^{0.2 \beta}=0.01 \pi 10^{0.2(3.47 \log P +K_0)},
$$ where $\pi$ is the parallax in mas. This allows us to use all the
parallax data, and thus minimize the bias which would be introduced by
selection. The right-hand side of the equation is weighted appropriately as
described by Whitelock \& Feast (2000a).

The equations were solved for various different subsets of the data, some of
which are listed in Table~1 (it may be necessary to add a bias correcion,
$\sim 0.04$\,mag to these values). The best result for the Miras is given by the
group of 180 stars which excludes the SP-red group and the small amplitude
variables, that should probably never have been considered as Mira-like. The
value of the zero-point for that group, $\beta=0.84\pm 0.14$, corresponds to
a distance modulus for the LMC of $18.64$, which is comparable to values
obtained in other ways. I should also draw your attention to the estimates
of the zero point for the SP-red and SP-blue stars listed in the table.
Given the large uncertainties the difference between them is not
significant, but this difference between the SP-reds and the main Mira group
(top row) is significant, and together with other evidence of differences,
suggests that the SP-red stars may be brighter than the Mira PL relation
would predict.
\begin{table}
\begin{center}
\caption[PL Zero-Point from the Hipparcos Parallax.]
{PL Zero-Point from the Hipparcos Parallax.}
\begin{tabular}{llll}
\sphline
No. Stars & $\beta$ (mag) & $\sigma_{\beta}$ & Stars Included\\
\sphline
180 & 0.84 & 0.14 & not SP-red; $\Delta Hp> 1.5$ mag\\
37 & 0.93 & 0.46 & SP-blue only\\
18 & 0.40 & 0.24 & SP-red only\\
38 & 0.90 & 0.31 & Carbon stars\\
\sphline
\end{tabular}
\vspace{-3mm}
\end{center}
\end{table}

It is worth making a comparison with the various sequences in the PL diagram
for the LMC, as discussed by Wood (2000) on the basis of Macho data. The
SP-blue stars are presumably identical to the normal Miras, i.e. sequence C
in Wood's nomenclature. According to Table~1 the SP-reds are about 0.5 mag
brighter than the SP-blues, whereas Wood's B sequence is about 1.3 mag
brighter. Note again that there is no difference in the mean pulsation
amplitude of the two groups.  It therefore seems that stars like these
SP-reds have not yet been identified in the LMC. The kinematic difference
between the SP-reds and SP-blues shows that they cannot be similar stars
pulsating in different modes. The possibility remains that the SP-reds have
a relationship to longer period Miras ($P>235$ days) which do have similar
kinematics. If that is the case then the SP-reds must be in a slightly
earlier evolutionary phase than Miras on the PL relation.

To conclude my comments on the local Miras, it would seem that,
with the exception of a small number with short periods and red colours,
the large amplitude variables fall on the same PL relation as do Miras
in the LMC and in globular clusters.

Finally, space has not permitted a discussion of the numerous OH/IR
variables near the galactic centre, which prevailing wisdom suggests are
faint, but which are difficult to measure accurately.  More work is also
need on C-star luminosities; the evidence points to their also obeying the
PL, but dust shells often make luminosity estimates difficult.

\begin{acknowledgments}
I wish to thank my various colleagues, particularly Michael Feast, 
Jacco van Loon, and Albert Zijlstra for their patience in the face of
my slow  progress towards publication.
My thanks also to Michael Feast and John Menzies for advice and critical 
reading of this manuscript.
\end{acknowledgments}

\end{document}